\documentclass[12pt]{report}
\usepackage{EMU_Thesis}
\usepackage{amsmath}
\usepackage{amssymb}
\usepackage{graphicx}
\usepackage{esint}
 \usepackage{pslatex}
\usepackage{maple2e}
\usepackage{graphics}
\DefineParaStyle{Maple Output}
\DefineParaStyle{Maple Output}
\DefineParaStyle{Text Output}
\DefineCharStyle{2D Math}
\DefineCharStyle{2D Output}

\title{Dark Matter; Modification of $f(R)$ or Wimps Miracle}
\degree{ Physics }
\author{Ali {\"O}vg{\"u}n}
\subyear{2013}
\Dept{Physics}

\InstituteDirector{Prof. Dr. Elvan Y{\i}lmaz}
\DeptChair{Prof. Dr. Mustafa Halilsoy}
\supervisor{Prof. Dr. Mustafa Halilsoy}
\cosuperi{Assist. Prof. Dr. Habib Mazharimousavi}
\examineri{Prof. Dr. Mustafa Halilsoy}
\examinerii{Assoc. Prof. Dr. \.{I}zzet Sakall{\i}}
\examineriii{Asst. Prof. Dr. Habib Mazharimousavi}
\dateofapproval{2013}

\begin{document}
\pagenumbering{roman} % Choose the relevant one
\makemstitle % For M.S. theses
%\makephdtitle      % For Ph.D. theses
%\makeproposaltitle % For Proposals

\makeapprovalpage

\begin{abstract}

\vspace{-2.0cm}

\noindent The identity of dark matter is one of the key outstanding problems in both
particle and astrophysics. In this thesis, I review some candidates of dark matter, especially WIMPs (weakly interacting massive particles) which is one of the best candidate so it is called that WIMPs miracle. In addition of this, there are also some theories of modification of gravity, by changing the law of gravity, it could be possible that the dark matter observations
are explained. Until the dark matter particle is detected, there is
some room for uncertainty. So we should consider every part of the problem and solve it. Dark matter problem is covering a large area so every possibility is important. So $f(R)$ gravity is also reviewed in this thesis and some theories are considered as a possible solution of dark matter problem. Finally we highlight that, even in the case of WIMPs or another particles solution, $f(R)$ gravity is also can be used for this problem. However, last words will be said by experiments.

\vspace{4.0cm}


\end{abstract}
% The usage of the "foreword" and "preface" environments are similar
% the "abstract" and "acknowledgements". See FBE manual for the
% correct order of these pages in the thesis.
\begin{ozet}

\vspace{-2.0cm}
\noindent Karanl{\i}k Maddenin gizemi hem astrofizik hem de par\c{c}ac{\i}k fizi\u{g}i a\c{c}{\i}s{\i}ndan \c{c}ok {\"o}nemli bir problemdir. Bu tezde, karanl{\i}k madde adaylar{\i}n{\i}, {\"o}zellikle WIMPs mucizesi olarak bilinen ve en g{\"u}\c{c}l{\"u} aday olan WIMPs(zay{\i}f etkile\c{s}en k{\"u}tleli par\c{c}ac{\i}k)i inceledim. Karanl{\i}k Maddenin gizemini \c{c}{\"o}zmeye \c{c}al{\i}\c{s}an bundan ba\c{s}ka teoriler de vard{\i}r, {\"o}rne\u{g}in gravitasyonun modifiye edilmesi gibi. Karanl{\i}k madde par\c{c}ac{\i}klar{\i} deneylerde bulunana kadar, bu gizem s{\"u}recektir. Biz bu gizemi her a\c{c}idan inceleyip \c{c}{\"o}z{\"u}m {\"u}retmemiz gerekir \c{c}{\"u}nk{\"u} \c{c}ok geni\c{s} bir alan{\i} kapsayan bu alan i\c{c}in her bir ihtimal bile \c{c}ok {\"o}nemlidir. Karanl{\i}k maddeyi a\c{c}{\i}klamaya aday olan $f(R)$ gravitasyon teorisini de bu tezde inceledik ve \c{c}{\"o}z{\"u}m {\"u}retmeye \c{c}al{\i}\c{s}t{\i}k. Sonu\c{c} olarak, gerek WIMPs olsun gerekse ba\c{s}ka par\c{c}ac{\i}klar olsun, $f(R)$ gravistasyon teorisi de g{\"u}zel bir alternatiftir ve kullan{\i}labilir. Tabii hangisinin do\u{g}ru oldu\u{g}unu deneyler onaylaycak.
\\
\\
\\
\\
\\

\textbf{Anahtar Kelimeler}: Karanl{\i}k Madde, Zay{\i}f Etkile\c{s}en K{\"u}tleli Par\c{c}ac{\i}klar(WIMPs), $f(R)$ Gravistasyon Teorisi, Kozmoloji.
 \end{ozet}

\newpage
\vspace*{11.5cm}
\centerline{To My Family}

\begin{acknowledgements} 
\vspace*{-.5cm}

\noindent I would like to express my deep gratitude to Prof.Dr.Mustafa Halilsoy, my supervisor, for his patient guidance, enthusiastic encouragement and useful critiques of this research work.

\noindent I would also like to thank Assoc.Prof.Dr.\.{I}zzet Sakall{\i}, for his advice and assistance in keeping my progress on schedule. 
My grateful thanks are also extended to Prof.Dr.{\"O}zay G{\"u}rtu\u{g}, Asst.Prof.Dr.Habib Mazharimousavi and  Asst.Prof.Mustafa R{\i}za.

\noindent I would also like to thank Pasquale Di Bari, who was my supervisor at University of Southampton for his advice on the Dark Matter production. 

\noindent I would like to thank my friends in the Department of Physics, The Gravity and General Relativity Group: K{\i}ymet Emral, Morteza Kerachian, Tayebeh Tahamtan,  Marzieh Parsa, Yashar Alizadeh for their support and for all the fun we have had during this great time. I wish to thank also other my friends for their support and encouragement throughout my study.

\noindent Finally, I would like to thank my family.

\end{acknowledgements}

\tableofcontents{}

\chapter{INTRODUCTION}
%\vspace{-10.5mm}
\pagenumbering{arabic} %\numberwithin{equation}{chapter}

\section{Observational Evidences for Dark Matter}
\vspace{-7.5mm}
\noindent Dark Matter is a very strange form of matter. It does not interact
with known objects because of the lack of electromagnetic interaction. However, it
has a mass and we know that it exists.The main reason for this view
is from gravitational interactions on the visible matter which holds
galaxies and galaxies of clusters together. The idea of the existence
of a new form of matter escaping electromagnetic detections, traces
back at least to 1915. The name ``dark matter'' was firstly used
by the Jacobus Kapteyn in 1922 in his studies of the motions of stars
in our galaxy \cite{1}. He found that there is no need of Dark Matter
within and around the Solar System. In 1923, Jan Oort arrived to an
opposite conclusion that there should be twice as much dark matter
as visible matter in the Solar vicinity. This is the first claim of
evidence for dark matter. However, since later observations have disproved
this early claim, in 1933 the discovery of dark matter is usually
credited to Fritz Zwicky who made the first correct claim about the
existence of dark matter.He concluded that the velocities of the galaxies
in the Coma cluster \cite{2} are much larger than the expected one.
Zwicky found that orbital velocities are almost a factor of ten larger
than expected from the total mass of the cluster \cite{29}. Zwicky
concluded that, in order to hold galaxies together in the cluster,
the cluster should contain huge amounts of some Dark (invisible) matter.
The galaxy rotation curve v, where r marks the distance from the
galaxy's center.

\begin{equation}
\frac{v^{2}}{r}=G\frac{M(r)}{r^{2}}\,,
\end{equation}

\noindent where M(r) is the mass inside radius r and $G_{N}=6.7\times10^{8}cm^{3}g^{-1}sec^{-2}$.
After rearranged the above equation reads 

\begin{equation}
v=\sqrt{\frac{GM(r)}{r}}\end{equation}

If the mass is only inside the radius, the velocity should decrease
when there is large value of radius as 
\begin{equation}
v\propto1/\sqrt{r}\end{equation} , where $G=1$.

However, for values of r bigger than the size of the galactic bulge,
the velocity of stars remains approximately constant\cite{22}. It means that there is a some unknown energy there such that
\begin{equation}
M(r)\propto r \end{equation}

\begin{figure}
\centering \includegraphics{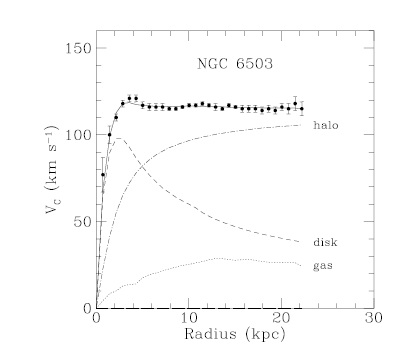} \caption[Distribution of dark matter in NGC 6503] \label{Distribution of dark matter in NGC 6503. Rotation curve of NGC 6503. The curves labeled ìgas,î ìdisk,î and ìhaloî represent the
contributions from the mass in gas, stars, and a dark halo. This plot of the rotation curve
is from Bertone et al. (2005b) using data from Begeman et al. (1991) \cite{22}}
\end{figure}

\noindent For an orbit around a compact central mass, such as planets in the
Solar system, we get Kepler's third law which is given at Eqn.(1.3).

\noindent As an example if the star is placed in the center of a galaxy, that is
the mass density of a galaxy decreases as a power law i.e
\begin{equation}
\rho\propto r^{-n}
\end{equation}
 The mass inside radius r is 
\begin{equation}
M(r)\propto\int r^{2}r^{-n}dr\propto r^{3-n}
\end{equation}

\noindent Thus the rotation velocity is 
\begin{equation}
v(r)=r^{1-n/2}
\end{equation}

\noindent Near the center of the galaxy, observed rotation curves of a galaxy
v(r) increase with r from Eqn.(1.3), however then suddenly becomes $v(r)\propto$
constant. From the Eqn.(1.5), the density profile is 
\begin{equation}
\rho\propto r^{-2}
\end{equation}

\noindent On the other side, the density of stars falls faster when goes to
the edges of the galaxy. The total mass from stars, gas and dust clouds, is
very very small so that they are not able to provide the rotational speed
at large distances. This means that there is another invisible, massive
matter. This matter would be only dominant in the outer parts. The
dark component forms a dark halo which holds galaxies and galaxies
of clusters together.

\begin{figure}
\centering \includegraphics{8} \caption[WMAP data] \label{WMAP data. Credit: NASA / WMAP Science Team \cite{25}}
\end{figure}

\noindent At galactic scales, there is also additional evidence for dark matter
which is based on the mass modeling of the detailed rotation curves
\cite{3}. An evidence of Dark Matter from cluster of galaxies was
first noticed, as mentioned already, by \cite{2} in 1933. On the cluster scales, the $\Omega_{M}$
is today`s mass density including baryonic mass and dark matter \cite{4,5}
 
\begin{equation}
\Omega_{M}\approx 0.2-0.3
\end{equation}

\begin{figure}
\centering \includegraphics{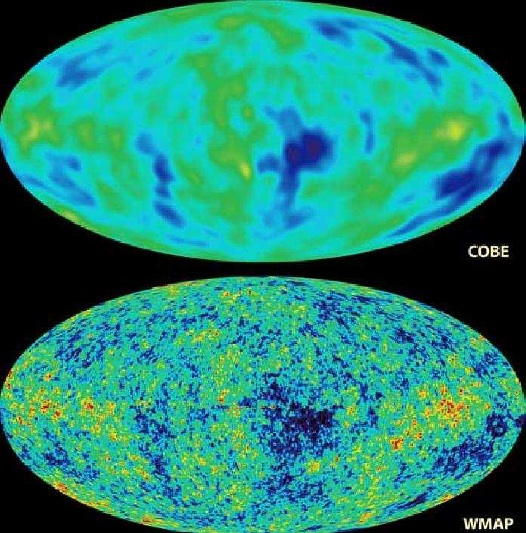} \caption[CMB Temperature fluctuations]\label{ CMB Temperature fluctuations. Image from NASA / WMAP Science Team \cite{25}}.
\end{figure}

\noindent As known that  matter relic density component is bounded by the observation
of the large scale structure.As reported by Sloan Digital Sky Survey
(SDSS) $\Omega_{M}h^{2}=0.135 $ for $\Omega_{b}h^{2}=0.025$
\cite{6}. The mass density of baryons and the mass density of matter in the universe
by using WMAP data are found to be \cite{18,7}  \\

\begin{equation}
\Omega_{b}h^{2}=0.0223
\end{equation}
\begin{equation}
\Omega_{M}h^{2}=0.127
\end{equation}

\begin{figure}
\centering \includegraphics[scale=0.75]{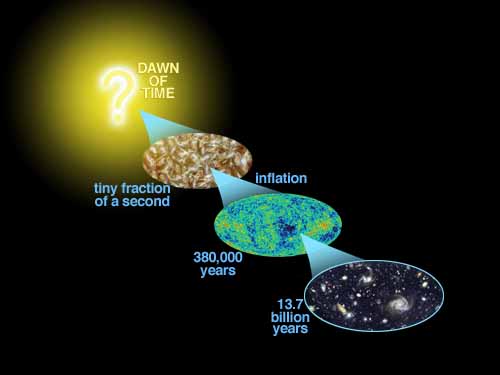} \caption[WMAP observes the first light of the universe]\label{WMAP observes the first light of the universe. Credit: NASA / WMAP Science Team \cite{25}} 
\end{figure}

\noindent From WMAP`s data and large scale structe observations`s data for relic abundance are consistent. In physical cosmology,
Big Bang nucleosynthesis provides us also a constraint\cite{13} 

\noindent The difference between $\Omega_{M}h^{2}$ and $\Omega_{b}h^{2}$ is
the clue of the existence of the non-baryonic dark matter.

\noindent There are many experiments which support the existence of Dark Matter
such as CMB anisotropy, galaxy surveys, etc. so we think that there
is Dark Matter. Unfortunately, the nature of dark matter is now an open problem.
So this is a very hot topic in physics.

\noindent Furthermore, there are some additional constraints on the amount of
dark matter in the Universe. We know very well from the concordance
cosmology measurements(it's basically a combination of looking at
the microwave background fluctuations, the baryon acoustic oscillations,
and the supernova brightness versus redshift relation)what are the
relative fractions of dark energy and mass in the Universe. We also
have pretty good constraints from primordial abundances of the elements
on what the relative fractions are for baryonic and non-baryonic dark
matter because if there were too high a density of protons in the
early universe, then there would have been more fusion in the early
universe\cite{30}.
\vspace{-7.5mm}

\section{Baryonic Dark Matter(BDM)}
\vspace{-7.5mm}
\noindent MACHO(massive compact halo object); planet-like objects, plasma and other compact objects are the candidates
for BDM. Other examples of this class of dark matter candidates include
primordial black holes created during the big bang, neutron stars,
white dwarf stars and various exotic stable configurations of quantum
fields, such as non-topological solitons. Gravitational Microlensing
is one of the ways to detect these objects. However, this method
only gives us the information about the mass, distance and velocity
of them and nothing more.
\\
\noindent Heavier, relatively faint objects such as old white dwarfs and mini
black holes,are the most important candidates for this BDM. If there
are primordial mini black holes produced in the big bang before BBN(Big Bang Nucleosynthesis),
they would not contribute to $\Omega_{b}$.
\\
\noindent $0.07M_{\odot}$ is required by a star to start to shine as a star
by igniting thermonuclear fusion. Brown dwarfs are not completely
dark because they radiate faint thermal radiation. However, the typical
mass of these MACHOs should be $\approx0.5M_{\odot}$, which is much
larger than the brown dwarf mass. In addition, white dwarf has a mass
which is suitable for dark matter but there is not enough number of
observed white dwarfs. Hence, BDM is not as good candidates for dark
matter as nonbaryonic dark matter.

\vspace{-7.5mm}
\section{Nonbaryonic Dark Matter}
\vspace{-7.5mm}
\noindent Hot Dark Matter (HDM) and Cold Dark
Matter (CDM) is considered as a nonbaryonic dark matter. Other more special options have been considered, such
as for example warm dark matter (WDM).\\

\noindent Dark matter particles that are called HDM, are relativistic at the time
of matter-radiation equality time. This requirement implies that their
mass has to be lower than about 100 eV. Their thermal velocities,
when structure formation begins are large because of the small mass.
In contrast, CDM has large mass so its velocities are negligible.
Candidates between hot and cold are called warm dark matter.\\

\noindent Effect of the pull of gravity causes to form galaxies and galaxy clusters
from early homogeneously distributed. This is called structure formation.
Today the best way to differentiate between HDM, WDM and CDM is through
the observed large-scale structure, i.e. CMB which shows the seeds
of structure. We show in Fig. (1.3) the results of a simulation of
the halo of dark matter around the Milky Way and two other galaxies.
For CDM, there is more substructure and satellites around the galaxy,
while their formation is suppressed for WDM. According to observations,
there is an order of magnitude less satellites observed around the
Milky Way than predicted by CDM models. However, the observations
are not complete, and the discrepancy may also be due to other causes
than WDM.

\begin{figure}
\centering \includegraphics[scale=0.5]{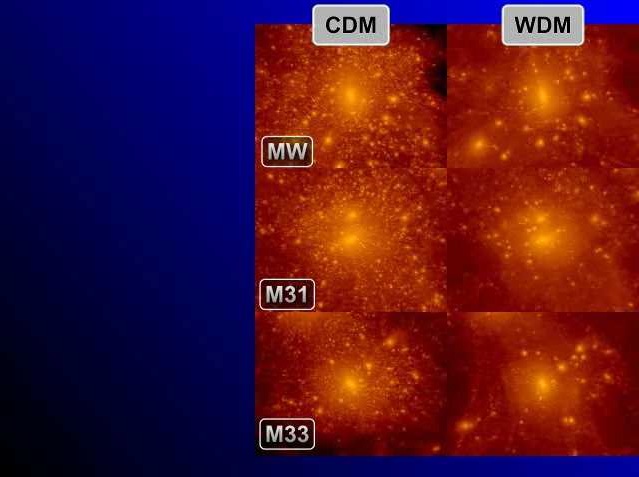} \caption[Comparison of the expected halo of the MilkyWay and the galaxies M31 and M33 in CDM and WDM models]\label{ Comparison of the expected halo of the MilkyWay and the galaxies M31 and M33 in CDM and WDM models. From http://www.clues-project.org/images/darkmatter.html.}
\end{figure}

\noindent The most common candidates for non-baryonic dark matter are Weakly
Interacting Massive Particles, or WIMPs \cite{17}. Cold dark matter has a most important candidate which is WIMPs. Because
they have neutral charge and they are long lived particles. Only weak
force and gravitational force effect them. Large amount of WIMPs are
produced and then the interaction rate of WIMPs was decreasing so
decoupling increasing between them in the thermal bath of the early universe when cools
down, like neutrinos, but are much heavier, so that they are a form
of CDM. After the decoupling, their relic density remains constant
as consistent with experiments. This is called WIMPs miracle. Furthermore
the interactions of some dark matter candidates are stronger or weaker.
For example, gravitinos have only gravitational interactions, while
TIMPs (Technicolour Interacting Massive Particles) have much stronger
interactions.
 \noindent Neutralino (bino+wino+2higgsino) lightest supersymmetric
particle is the best candidate for WIMP in Minimal Supersymmetric Standard Model (MSSM)\cite{14,15}.
\noindent The other candidate for WIMP is the lightest Kaluza Klein (LKP) which
is coming from Universal extra-dimensions (UED) model .

\vspace{-7.5mm}
\subsection{Hot Dark Matter (HDM)}
\vspace{-7.5mm}
\noindent For the HDM neutrinos are candidates. Energy density today is dominated
by their rest masses so to make a significant contribution to the
density parameter, neutrinos should have a rest mass about 1 eV.
\noindent The present upper limit to HDM in the form of massive neutrinos leads
to the conservative mass limit from WMAP data.

\begin{equation}
\sum m_{\nu}\leq0.62eV
\end{equation}

\noindent Therefore the maximum contribution of neutrinos is $\Omega_{\nu}h^{2}\leq0.02$,
at most the same order of magnitude as baryonic matter. If neutrinos
were the dominant form of matter, there would be a lower limit on
their mass from constraints on the phase space density, called the
Tremaine-Gunn limit. Essentially, in order to achieve a certain rotation
velocity for galaxies, you need a certain amount of mass inside a
given volume, and the Pauli exclusion principle constrains the number of particles you can pack inside a given volume. Even though
we know that neutrinos are a subdominant component of dark matter,
the Tremaine-Gunn limit applies to any fermionic dark matter candidate,
even if its distribution is not thermal. (There is no similar lower
limit on the mass of a bosonic dark matter particle.)

\vspace{-7.5mm}
\subsection{Cold Dark Matter (CDM)}
\vspace{-7.5mm}
\noindent Most of the matter in the universe has to be CDM because of data on
large scale structure combined with structure formation theory. WIMPs
(Weakly Interacting Massive Particles) are the main candidates of
CDM.
\noindent Dark matter candidates are with masses in the 100 GeV to TeV range.
Particles with GeV to TeV masses and weak-scale interaction cross
sections are known as Weakly Interacting Massive Particles (WIMPs).
There are also another candidates for DM. However, WIMP dark matter
is compelling for two main reasons:
\begin{enumerate}
\item In the early universe, when WIMPs were in thermal equilibrium annihilated
with each other with a weak scale cross section. As the universe expanded
and has frozen out, WIMPs cooled to obtain a relic density today which
is equal to the measured dark matter density.

As mentioned before that this is known as the WIMP Miracle.

\item The other reason is that WIMP dark matter are present in many of the
favored extensions of the Standard Model.For example, the Minimal
Supersymmetric Standard Model (MSSM) has a favourite WIMP candidate. 
\end{enumerate}

\chapter{EXTENDED GRAVITY: $f(R)$ GRAVITY}

\section{Introduction to $f(R)$ Gravity}
\vspace{-7.5mm}
\noindent By changing the law of gravity, it could be possible that the observations
are explained. Until the dark matter particle is detected, there is
some room for uncertainty. The problem for modified gravity is that
there are so many different observations for dark matter, in different
physical systems: motions of stars in galaxies, motions of galaxies
in clusters, gravitational lensing, large-scale structure, CMB anisotropies
and so on. Gravity has to be adjusted in a different manner for these
different observations, and the resulting models are somewhat contrived.
Expressed another way, the dark matter scenario is very predictive:
the simple hypothesis of a massive particle with weak couplings to
itself and to the Standard Model particles explains a number of disparate
observations and has made successful predictions.\\
\noindent In 1919 and 1922 respectively by Weyl and Eddington \cite{51,52} 
gravitational action firstly studied. However, until 1960s, there
was not any indications of complicating the gravitational action.
Because of cosmological problems such as the dark energy problem,
dark matter problem, this type of modification is needed to fix the
problems. At $f(R)$ theory, the main idea is to take action as a function
instead of a constant curvature such as:

\begin{equation}
S=\frac{1}{16 \pi G}\int d^{4}x\sqrt{-g}[f(R)]
\end{equation}

\noindent By using the $f(R)$ function , it provides usage of many different properties,
which are very important for explaining unknown matters
such as dark matter and dark energy. For instance, considering f as
a series expansion i.e \cite{54,55,56,57,58,59} 

\begin{equation}
f(R)=...+\frac{\alpha_{2}}{R^{2}}+\frac{\alpha_{1}}{R}-2\Lambda+R+\frac{R^{2}}{\beta_{2}}+\frac{R^{3}}{\beta_{3}}
\end{equation}
\noindent where $\alpha_{i}$ and $\beta_{j}$ coefficients have the appropriate
dimensions. So it can be seen that a number of interesting terms appear. \cite{53}

\noindent Without matter action is given above, now after adding matter on the
action, our $f(R)$ action is

\begin{equation}
S=\frac{1}{16\pi G}\int d^{4}x\sqrt{-g}[f(R)+S_{M}(g_{\mu\nu},\psi)].
\end{equation}

\noindent Varying the action with respect to $g_{\mu\nu}$ provides

\begin{equation}
f_{R}R_{\mu\nu}-\frac{1}{2}fg_{\mu\nu}-\nabla_{\mu}\nabla_{\nu}f_{R}+g_{\mu
\nu}\square f_{R}=\kappa T_{\mu\nu}%
\end{equation}

\noindent where $f_{R}=\frac{df}{dR}$ \ , $\square f_{R}=\frac{1}{\sqrt{-g}}%
\partial_{\mu}\left(  \sqrt{-g}\partial^{\mu}\left(  \frac{df}{dR}\right)
\right)  $ \ \ and \ $\nabla^{\nu}\nabla_{\mu}f_{R}=g^{\alpha\nu}\left[
\left(  f_{R}\right)  _{,\mu,\alpha}-\Gamma_{\mu\alpha}^{m}\left(
f_{R}\right)  _{,m}\right]  $

So

\begin{equation}
\square f_{R}=\square\frac{df}{dR}=\frac{1}{\sqrt{-g}}\partial_{1}\left(
\sqrt{-g}g^{11}\partial_{1}\left(  \frac{df}{dR}\right)  \right)  =\frac
{1}{\sqrt{-g}}\partial_{1}\left(  \sqrt{-g}g^{11}\left(  \frac{df}{dR}\right)
^{^{\prime}}\right)
\end{equation}

\begin{equation}
\nabla^{0}\nabla_{0}f_{R}=\nabla^{0}\nabla_{0}\frac{df}{dR}=-\frac{1}{2}%
g^{00}\Gamma_{00}^{1}\left(  \frac{df}{dR}\right)  ^{^{\prime}}=-\frac{1}%
{2}g^{00}g^{11}g_{00,1}\left(  \frac{df}{dR}\right)  ^{^{\prime}}%
\end{equation}

\begin{equation}
\nabla^{1}\nabla_{1}f_{R}=\nabla^{1}\nabla_{1}\frac{df}{dR}=g^{11}\nabla
_{1}\nabla_{1}\frac{df}{dR}=g^{11}\left(  \frac{df}{dR}\right)  ^{^{^{\prime
\prime}}}-g^{11}\Gamma_{11}^{1}\left(  \frac{df}{dR}\right)  ^{^{\prime}}%
\end{equation}

\begin{equation}
\nabla^{2}\nabla_{2}f_{R}=-g^{22}\Gamma_{22}^{1}\left(  \frac{df}{dR}\right)
^{^{\prime}}=\frac{1}{2}g^{22}g^{11}g_{22,1}\left(  \frac{df}{dR}\right)
^{^{\prime}}%
\end{equation}

\noindent in which a prime means d/dr and its variation with respect to $g_{\mu\nu}$ yields, after some
manipulations and modulo surface terms, the field equation

\begin{equation}f_{R}R_{\mu\nu}-\frac{f(R)}{2}g_{\mu\nu}=\nabla_{\mu}\nabla_{\nu}f_{R}-g_{\mu\nu}\Box f_{R}+\kappa T_{\mu\nu}\end{equation}

\noindent where $\nabla_{\mu}$ is the covariant derivative associated with
Levi-Civita connection of the metric, and $\Box=\nabla^{\mu}\nabla_{\mu}$.

\noindent These equations above are fourth order partial differential equations
in the metric. The alternative name ``fourth order gravity'' used
for this theories because in the first two terms on the right hand
side fourth order derivatives appear.

\noindent Furthermore, the contraction of the  above equation  yield the
trace equation
\begin{equation}
f_{R}R-2f+3\square f_{R}=\kappa T
\end{equation}

\noindent where $T=T_{\mu}^{\mu}=0$

\noindent the trace equation can be used to simplify the field
equations and then keep it as a constraint equation.

\noindent Therefore
\begin{equation}
f_{R}R_{\nu}^{\mu}-\frac{1}{4}\delta_{\nu}^{\mu}\left(  f_{R}R-\square
f_{R}-\kappa T\right)  -\nabla^{\nu}\nabla_{\nu}f_{R}=\kappa T_{\nu}^{\mu}%
\end{equation}

\noindent The trace of this equation is

\begin{equation}
f_{R}R-2f(R)+3\Box f_{R}=8\pi GT
\end{equation}
where $T=g^{\mu\nu}T_{\mu\nu}$

\noindent Here we can see that when $T=0$ , not anymore says that $R=0$ or constant.

\noindent Another important thing is that when $R=constant$ and $T_{\mu\nu}=0$,
it reduces to

\begin{equation}f_{R}R-2f(R)=0\end{equation}

\noindent Maximally symmetric solutions is affected by f function, and they
will be Minkowski, de Sitter or anti-de Sitter.
Hence, We can now write the field equations in the form of Einstein
equations as

\begin{equation} G_{\mu\nu}=R_{\mu\nu}-\frac{1}{2}g_{\mu\nu}R=\frac{\kappa T_{\mu\nu}}{f_{R}}+g_{\mu\nu}\frac{[f(R)-Rf_{R}]}{2f_{R}}+\frac{[\nabla_{\mu}\nabla^{\nu}f_{R}-g_{\mu\nu}\Box f_{R}]}{f_{R}}   \end{equation}

\noindent After some algebra the above equation, takes the form

\begin{equation} F(R)R_{\mu\nu}-\nabla_{\mu}\nabla^{\nu}F(R)-kT_{\mu\nu}=-g_{\mu\nu}(\frac{1}{2}f(R)-\nabla_{\alpha}\nabla^{\alpha}F(R))\end{equation}  where $F(R)=df/dR=f_{R}$.

\noindent For solving above equation we need a metric and its components\cite{53}
. Here we are using spherically symmetric metric with radial components B(r) and X(r) :

\begin{equation} ds^2=-B(r)dt^2 + \frac{X(r)}{B(r)}dr^2 + r^2d\Omega^2 \end{equation}

\noindent  Ricci scalar of this metric can be written as 

\begin{equation} R(r)=\frac{2}{r^2}+\frac{X'} {X^2}(\frac{B'}{2}+\frac{2B}{r})-\frac{1}{X}(B''+\frac{4B'}{r}+\frac{2B}{r^2}) \end{equation}

\noindent For empty space if we solve the Einstein Equation with the form of $f(R)$ gravity, two independent field equations are found :

\begin{equation} \frac{X'}{X} = \frac{2rF''}{2F+rF'}  \end{equation}

\begin{equation} B''+(\frac{F'}{F}-\frac{X'}{2X})B'-\frac{2}{r}(\frac{F'}{F}-\frac{X'}{2X})B-\frac{2}{r^2}X=0 \end{equation}

\noindent where the prime means d/dr. After that are choose our ansatz with constants $ F_{0}$, $X_{0}$ and $B_{0} $ as 

\begin{equation} F(r)=F_{0}r^{n}, X(r)=X_{0}r^{\alpha}, B(r)=B_{0}r^{\alpha} \end{equation}

\noindent Constans are bounded by two constraints as:

\begin{equation} \alpha=\frac{2n(n-1)}{n+2}, X_{0}=[1+\frac{n(2-\alpha)}{2}-\frac{\alpha^2}{4}]B_{0}.\end{equation}

\noindent After that \begin{equation} R(r)=\frac{2R_{0}(n,\alpha)}{r^2} \end{equation} is obtained.

\noindent where

\begin{equation} R_{0}(n,\alpha)=1-\frac{1+\frac{\alpha}{2} +\frac{\alpha^2}{8}}{1+n(1-\frac{\alpha}{2})-\frac{\alpha^2}{4}} \end{equation} 

\noindent This will be the well known solution of Schwarzschild  when $\alpha$ or n goes to zero. After using those constraints and definitions, an asymptotic form of $f(R)$ where 
$\Lambda_{0}$= constant of integration is found as 

\begin{equation} f(R)= \frac{f_{0}(2R_{0})^{(\frac{n}{2})}}{1-(\frac{n}{2})}R^{(1-\frac{n}{2})} + \Lambda_{0} \end{equation}

\noindent In the paper of A Model of $f(R)$ Gravity as an Alternative for Dark Matter in Spiral Galaxies they took $F_{0}=1 $ when n is going to zero and action goes to $F_{0}R+\Lambda_{0}$.

\noindent For the application  of this in the rotation curve, geodesic equation is used in the weak field approximation. The test particle is rotating around the central mass and obtains as 
\begin{equation} \ddot{r}+\Gamma^{r}_{tt}=0 \end{equation}  where the dot means d/dt. After using the affine connection which we found by using metric :

\begin{equation}
\Gamma_{t t}^r = \frac{B}{2XB'} , \Gamma_{t r}^t = \frac{B'}{2B} , \Gamma_{r r}^r = \frac{X'B-XB'}{2BX} ,  \Gamma_{r \theta}^{\theta} = \Gamma_{r \phi}^{\phi}=\frac{1}{r} ,
\Gamma_{\theta \theta}^r = \frac{Br}{X} \end{equation}

\begin{equation}
 \Gamma_{\theta \theta}^r = \frac{Br}{X} , \Gamma_{\theta \phi}^{\phi} = tan(\theta) , \Gamma_{\phi \phi}^{r} =  \frac{Br sin(\theta)^2}{X} , \Gamma_{\phi \phi}^{\theta} = -\frac{sin(\theta)}{2} 
\end{equation}

\noindent velocity for a test particle is obtained as 

\begin{equation} v^{2}=\frac{rc^{2}B'B}{X}= \frac{c^{2} \alpha B_{0}r^{\alpha}}{2} \end{equation}

\noindent After considering some assumption for the large distances from center of galaxy, such as using  $r^{\alpha}=1+\alpha\ln(r)$  where $\alpha$ is  smaller than 1, one finds that

\begin{equation}v^{2}_{\infty}=\frac{c^{2}\alpha B_{0}}{2}  \end{equation}

\noindent For taking \(B_{0}=2\sqrt{\mu M_{galaxy}}\) where $\mu$ is the proportional coefficient with mass inverse dimension and M is mass of galaxy, hence asymptotic velocity is found

\begin{equation} v^{2}_{\infty}=c^{2}\alpha\sqrt{\mu M_{galaxy}}  \end{equation} 

\noindent where $\alpha$ and $\mu$ depend on experimental datas. For instance, if this values taken as  \( \alpha^{2} \mu=(v_{\infty}/c)^{4} /M_{galaxy}=10^{-54}kg^{-1}\), the velocity is found exactly $200 kms^{-1}$ and this value is consistent with Tully-Fisher`s relation. \cite{53}

\noindent Furthermore,  another good testing is made by using the Modified Newtonian dynamics (MOND).

\begin{equation} \alpha^{2}\mu=0.98\times10^{-54}kg^{-1} \end{equation} for the velocity of typical galaxy.

 \noindent There are many difficulties of this theory. One example which gained a lot of publicity a few years ago is the
Bullet cluster. A solution of the Dark matter conundrum in terms of a new elemantary particle is quite successful to explain the Bullet cluster, while explanations in terms of modified gravity models encounter serious difficulties. Another problem in this way of solution is that ansatz is choosen by hand and not naturally comes out.

% Chapter 2

\chapter{DARK MATTER PRODUCTION}

% Write in your own chapter title

\section{Thermal Production}
\vspace{-7.5mm}
\noindent WIMPs and many other dark matter candidates are in thermal equilibrium
in the early universe and decouple once their interactions become
too weak to keep them in equilibrium. Those particles are called thermal
relics, as their density today is determined by the thermal equilibrium
of the early universe. When the annihilation processes of WIMPs into
SM particles and vice versa happend at equal rates, there is a equilibrium
abundance. When the Universe cooled down and the rate of expansion
of the universe H exceeds the annihilation rate, the WIMPs effectively
decouple from the remaining SM particles. So the equilibrium abundance
drops exponentially until the freeze-out point.The abundance of cosmological
relics remained almost constant until today \cite{35}.

\noindent If the candidate is stable or has a long lifetime, the number of particles
is conserved after it decouples, so the number density falls like
$a^{-3}$. Specifically, we use the Lee-Weinberg Equation \cite{61}. It describes
annihilation and creation of $\chi$ particles \cite{60,62}.

\begin{equation}
\frac{dn_{\chi}}{dt}+3Hn_{\chi}=-<\sigma v>(n_{\chi}^{2}-n_{\chi,eqn}^{2})
\end{equation}

\noindent with $n_{\chi,eqn}^{2}$ ,the  number density of the
relic particles in equilibrium. Furthermore Hubble constant make expanding universe effect. The thermal average of the
annihilation cross section $\sigma$ multiplied with the relative
velocity v of the two annihilating $\chi$ particles. The first (second)
term on the right-hand side (RHS) of the equation describes the decrease
(increase) of the number density due to annihilation into (production
from) lighter particles. Kinetic equilibrium  is satisfied by the Lee-Weinberg equation for $\chi$

\noindent Now we use Hubble constant effect which provide  universe expansion by considering
the evolution of the number of particles in a portion of comoving
volume given by

\begin{equation}
N=n_{\chi}R^{3}
\end{equation}
We can then introduce the convenient quantity 
\begin{equation}
\bar{N}=\frac{n_{\chi}R^{3}}{N(0)}
\end{equation}

\noindent such that $\bar{N(0)}=1$.

\noindent In addition, since the interaction term usually depend explicitly
upon temperature rather than time, we introduce the $x=\frac{m}{T}$, the
scaled inverse temperature.

\noindent During the radiation dominated period of the universe, thermal production
of WIMPs takes also place . For this time, one can find the expansion rate

\begin{equation}
H=\sqrt{\frac{8\Pi^{3}gg_{\chi}T^{2}}{90}}
\end{equation}
and then 
\begin{equation}
\Longrightarrow H(x=1)=m^{2}\sqrt{\frac{8\pi^{3}gg_{\chi}}{90}}
\end{equation}
where $H(x)=\frac{H(x=1)}{x^{2}}$ and where $_{\chi}$ refer the
number of intrinsic degrees of freedom such a like that spin or color for $\chi$ particle.

\noindent Hence we can write the last format of our equation a

\begin{equation}
\frac{d\bar{N}}{dt}=-\frac{<\sigma v>N(0)}{R^{3}Hx}(\bar{N}^{2}-\bar{N}_{eqn}^{2})
\end{equation}
where 
\begin{equation}
\frac{N(0)}{R^{3}}=\frac{3\zeta[3]m^{3}}{4 \Pi^{2} x^{3}}
\end{equation}

\noindent After all little tricks , our Lee-Weinberg equation can be recast
as

\begin{equation}
\frac{d\bar{N}}{dt}=-\frac{\Psi(\chi)}{x^{2}}(\bar{N}^{2}-\bar{N}_{eqn}^{2})
\end{equation}

\noindent where we defined 
\begin{equation}
\Psi(\chi)=\frac{3\zeta[3]}{4\Pi^{2}}\sqrt{\frac{45}{4\Pi^{3}}}\frac{g_{\chi}}{\sqrt{g_{\chi}}}mM_{Pl}<\sigma v>=0.055\frac{g_{\chi}}{\sqrt{g_{\chi}}}mM_{Pl}\sigma_{0}
\end{equation}

where $\zeta$ is EulerñRiemann zeta function
\noindent To integrate the Lee-Weinberg equation, we need to have an expression
for the equilibrium number density in comoving volume.Once the particle
is non-relativistic, the difference in statics is not important.

\noindent The general equation of equilibrium number density in comoving volume
is

\begin{equation}
\bar{N(x)}_{eqn}=\frac{2}{3\zeta[3]}\int_{0}^{\infty}\frac{y^{2}}{1+\exp^{\sqrt{x^{2}+y^{2}}}}dy
\end{equation}

\noindent For the nonrelativistic case at low temperatures $T\gg m_{\chi}$ one
obtains, 
\begin{equation}
\bar{N(x)}_{eqn,NR}=\frac{2}{3\zeta[3]}\frac{\Pi}{2}x^{3/2}\exp^{-x}
\end{equation}

\noindent At great value of temperatures, dark matter quickly is destroyed by its own antiparticle. Shortly after that, until $\Gamma_{\chi}=\bar{N(x)}<\sigma v> \ll  H $, T has dropped too much which is under the value of  $m_{\chi}(T<<m_{\chi})$, also the number density of dark matter drops exponentially. The temperature at which the particle decouples (The time when the number of particles reaches this constant value) is called freeze-out temperature $T_{F}$. Hence the number density per comoving volume becomes almost constant because dark matter particles are no longer able to annihilate efficiently.

\noindent An approximate solution for the relic abundance is given by 
\begin{equation}
\bar{N}(t)\equiv\bar{N}(\infty)\cong\frac{x_{f}}{\Psi}
\end{equation}
. where $x_{f}=\frac{m}{T_{f}}$ and $T_{f}$ is the so called `freeze-out`
temperature and approximately, for the typical case $x_{f}\succ\succ1$,one
has that $x_{f}$ is a solution of

\begin{equation}
\sqrt{x_{f}}e^{x_{f}}=\Psi
\end{equation}

\noindent We can then finally calculate the contribution of $\chi$ to the energy
density parameter finding a well know result

\begin{equation}
\Omega_{\chi}h^{2}\cong4\times10^{-10}\frac{GeV^{-2}}{\sigma_{0}}
\end{equation}

\noindent It is intriguing that so called ``WIMPS''(e.g. the lightest supersymmetric
particle) dark matter particles seem to reproduce naturally the right
abundance since they have a weak cross section

\begin{equation}
\sigma_{0}\cong\frac{\alpha^{2}}{m^{2}}\cong10^{-9}(\frac{100GeV}{m^{2}})
\end{equation}
and masses $m_{x}~m_{ew}~100GeV$. This observations is called the
``WIMP miracle'' and typically considered as an encouraging point
supporting WIMPS as Dark Matter candidates.
\begin{figure}
\centering \includegraphics{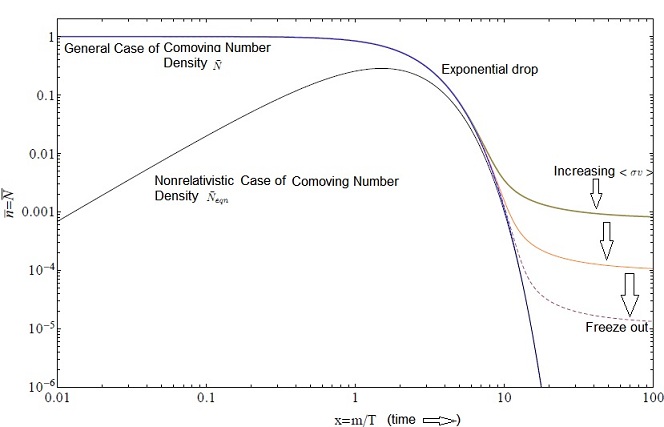} \caption[ Evolution of $\bar{N}$  and $\bar{N}_{eqn} $  as function of x ] \label{ Evolution of $\bar{N}$  and $\bar{N}_{eqn} $  as function of x for m =100 GeV, $g_{\chi} =2 $, g =90 (constant), $\sigma_{0}=10^{-14} , 10^{-15} and  10^{-16} $  , $M_{Pl} =1.9 \times10^{19}  GeV^2 $   }
\end{figure}

\noindent The LW eqn. can be solved by approximately analytic
calculation or numerical plotting by using Mathematica.We have plotted
those equation by using both of the ways. As you can see Fig. 2.1.

\noindent LW Eqn., the termal production, is the most traditional
mechanism but that many other mechanisms have been considered in the
literature such as non thermal production of very massive particles(so
called WIMPZILLAS) at preheating or even right-handed neutrino oscillations.

% Chapter 6

\chapter{CONCLUSION}

\noindent The observational evidence for dark matter is overwhelming and now
we are entering one of the perhaps most exciting periods in physics
history where a wide array of experiments may have the potential to
unveil its nature.In this thesis I review some important concepts
about dark matter. One of the best motivated candidates for dark matter
so far is the WIMPS which I mentioned before. In this thesis the focus
was on dwarf spheroidal galaxies, promising targets that are assumed
to be strongly dark matter dominated systems due to their high mass-tolight
ratios up to few hundred solar units. Dwarf galaxies of the Milky
Way have the advantage of relative proximity and low astrophysical
background as they do not lie within the galactic plane. Great importance
was given to studying the properties of these systems and how their
dark matter profiles can be derived from kinematical studies. However
a precise estimation of the density distribution is a challenging
task and due to too little data available in particular the shape
of the dark matter halo inner regions remains poorly known. Cusped
profiles suggested from simulations like the Navarro-Frenk-White profile
are not ruled out in general although observations seem to favor profiles
with a flat inner core.

\noindent The two investigated dwarf spheroidal galaxies Sagittarius and Canis
Major are particularly attractive targets as they are the closest
dwarf galaxies discovered so far. The uncertainty concerning their
inner region is reflected in the plots of sensitivity limits for both
investigated dwarf galaxies, where the distances between the two lines
of a cored and a cusped profile are about one order of magnitude.

\noindent Better opportunities for a significant dark matter detection could
be obtained for instance by the next generation ground-based CTA (Cherenkov
Telescope Array) experiment with bigger effective areas due to many
Cherenkov telescopes in the order of 50 and also IceCube experiment
. Here, stereoscopy will be possible for small energies. Moreover
the space based ray telescope FERMI will yield interesting results
in the future.Apart from Sagittarius and Canis Major, there could
be promising faint dwarf galaxies that are not discovered yet. However,
there should be as many undiscovered dwarf galaxies in the other direction,
too, which, when discovered, will provide interesting targets for
the search of dark matter.

\noindent In chapter 2 we gave a simple example of $f(R)$ extended gravity in which the effect of DM is not due to exotic sources but due to the wrong choice of Lagrangian. Einstein`s model, which extends that of Newton is characterized by the specific choice $f(R)=R$. In the $f(R)$ model this has been modified to different polynomial powers of R. As an example the choice $f(R) \approx R^{1-\frac{n}{2}}$ where $ n\ll 1 $, describes asymptotic behaviours of flat rotation curves and experimental results of Tully-Fisher (i.e. $v^{4}\approx$ Mass). For n=0 the model reduces to the Einstein`s general relativity model which fails to explain flat rotation curves unless supplemented by exotic sources.

% You can also try harvardbibliography, bibnotcited and
% harvardbibnotcited environments for other types of bibliographical
% lists.

\end{document}